# Entangled *N*-photon states for fair and optimal social decision making

Nicolas Chauvet[1*], Guillaume Bachelier[2], Serge Huant[2], Hayato Saigo[3], Hirokazu Hori[4], Makoto Naruse[1]

[1] Department of Information Physics and Computing, Graduate School of Information Science and Technology, The University of Tokyo, 7-3-1 Hongo, Bunkyo-ku, Tokyo 113-8656, Japan

[2] Univ. Grenoble Alpes, CNRS, Institut Néel, 38000 Grenoble, France

[3] Nagahama Institute of Bio-Science and Technology, 1266 Tamura, Nagahama, Shiga 526-0829, Japan

[4] Interdisciplinary Graduate School, University of Yamanashi, Takeda, Kofu, Yamanashi 400-8510, Japan

* nicolas_chauvet@ipc.i.u-tokyo.ac.jp

**Abstract**

**Situations involving competition for resources among entities can be modeled by the competitive multi-armed bandit (CMAB) problem, which relates to social issues such as maximizing the total outcome and achieving the fairest resource repartition among individuals. In these respects, the intrinsic randomness and global properties of quantum states provide ideal tools for obtaining optimal solutions to this problem. Based on the previous study of the CMAB problem in the two-arm, two-player case, this paper presents the theoretical principles necessary to find polarization-entangled *N*-photon states that can optimize the total resource output while ensuring equality among players. These principles were applied to two-, three-, four-, and five-player cases by using numerical simulations to reproduce realistic configurations and find the best strategies to overcome potential misalignment between the polarization measurement systems of the players. Although a general formula for the *N*-player case is not presented here, general derivation rules and a verification algorithm are proposed. This report demonstrates the potential usability of quantum states in collective decision making with**

limited, probabilistic resources, which could serve as a first step toward quantum-based resource allocation systems.

INTRODUCTION

Decision making research is quite diverse and involves the study of optimal algorithms and to make the best decisions possible in particular situations, with applications ranging from psychology[1-3] and management[4-6] to reinforcement learning[7] and dynamic resource allocation[8-10]. Fundamentally, this type of research addresses how complete or partial information about a given environment is processed by a "user" (human or artificial) to achieve the best possible outcome.

The multi-arm bandit (MAB) problem is a typical example of decision making inspired by game theory, in which a user faces several choices with identical potential reward amounts (slot machines in the canonical example), whose reward probabilities are unknown to the user but can change without notice. Based on real-life situations such as mobile network connections[11], computing[10], or industrial process chains[8], the method that the user typically adopts is to try each machine and select the one with the highest reward probability, while still trying the others from time to time to verify whether or not the current choice remains the best.

In many application instances, another dimension must be accounted for: the competition for rewards/resources between multiple users choosing simultaneously, such as for traffic regulation, telecom bandwidth allocation, or energy grid management. In these cases, users should not only search for the best choice, but also ensure that the reward allocation is both globally optimal and fair for all users. When the number of choices is limited and the number of users is high, the main focus of the problem becomes ensuring that the reward is equally shared rather than finding the best choice individually, such as in the case of mobile network connections in crowded environments.

For the MAB problem, previous researchers have implemented situations by using photonic systems as a resource for decision making, including chaotic laser sources[12,13], excitation transfer via near-field coupling[14,15], or polarized single-photon sources[16,17]. Using already established algorithms[18], these works have demonstrated the potential usability of physics to solve complex decision-making situations efficiently, in up to a 64-arm single-user case[13].

A consistent extension of this work is to study situations in which several players must make choices simultaneously between multiple unknown, probabilistic choices, known as the competitive multi-arm bandit (CMAB) problem. Instead of using parallel, independent physical systems to solve this problem, another idea is to use particular collective states following quantum formalism, known as entangled states, to study collective strategies. Entanglement is a fundamental property of composite systems in quantum physics that has attracted interest in game theory for solving deterministic problems with payoff matrices by finding Nash equilibrium[19] in competitive situations[20-22]. Some reports have also highlighted the benefits of using entanglement and a quantum approach in machine learning[23-25]. However, as the CMAB problem is probabilistic and the same kinds of solution algorithms cannot be followed, new approaches are needed.

We previously numerically and experimentally demonstrated that the polarization degree of freedom of photons can be used to allocate the rewards from two machines to two users efficiently by using polarization-entangled photon pair states[26]. In the present report, we extended this work to the use of $N$ polarized photons in an $N$-photon quantum superposition state to solve the CMAB in more general situations with $N \geq 2$ and two choices. The goal is to maximize both the reward output from the machines and the equality, or fairness, of the repartition of rewards among players, given that their possible actions are limited to a rotation of their own polarization measurement basis at every turn. More precisely, the intent is to obtain $N$-photon quantum superposition states with such properties that the optimal strategy for each player leads to an optimal situation for all players in terms of total outcome and fair repartition.

In this report, after defining the general situation under study, we identify the set of constraints on $N$-photon quantum superposition states to optimize the reward allocation for any $N$. From there, we present the derivation of the corresponding states for three, four, and five users and a comparison of the corresponding properties for the MAB solution. Finally, we discuss the convergence estimation of error-correction protocols with these states.

## RESULTS

### Definitions and formulation of the problem

**Problem description and hypotheses**

The MAB problem is used to reproduce situations in which the available choices have outcomes with different probabilities and/or reward amounts, which are unknown to the user(s). In the CMAB problem, several users simultaneously face the same choices, also referred to as "machines" in the following, with or without communication with each other. If several users select the same machine, the potential reward is split between them. Figure 1 represents a typical example case of internet connection through relays, where the concepts of competition and limited communication between users are clearly involved.

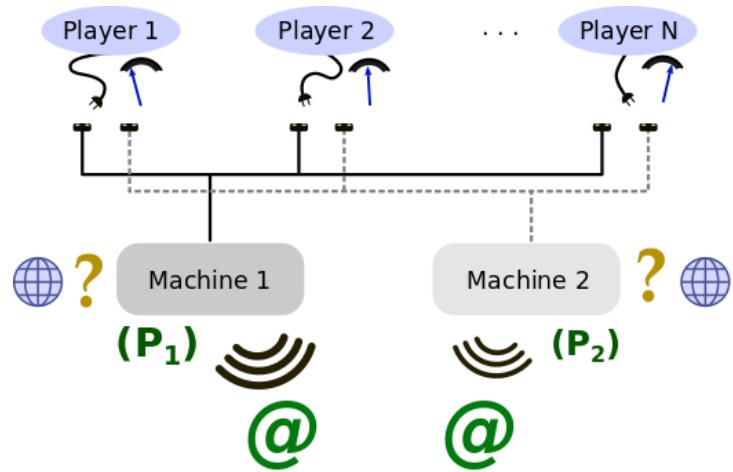

**Figure 1**. Principle of CMAB problem with $N$ players and two machines, illustrated with internet connection. In every run, each player selects one machine to use to try to connect to the internet, which does so with a (unknown) probability $P_{1,2}$. In case of connection, each player receives the same, shared bandwidth as the other players who selected the same machine.

According to the situation in Fig. 1, we defined four rules for our study:

(i) Each machine has a fixed reward in case of success, equal to 1 for all, and 0 otherwise;

(ii) If a machine is selected by *k* users and gives a reward, each user receives an individual reward of 1/*k*;

(iii) The users know the reward amount given by each machine in case of success, as well as the total number of users, but not the probability of reward of the machines;

(iv) The users do not communicate with each other.

Rule (i) is only for calculational convenience and does not cause loss of generality, as arbitrary and different rewards between machines produce the same results. The main requirements for Rule (ii) are that the total rewards of the machines remain the same regardless of how many users select them, and that some users are not favored over others. Without Rule (iii), users would have no means of deducing anything about the decisions of the other users and their consequences; thus, it is necessary to elaborate the optimization algorithms. Rule (iv) restricts the kinds of algorithms usable here and emulates actual network architecture constraints.

**Definitions and formalism**

We consider now the CMAB with $N \geq 2$ and two machines, A and B. Let $x_{i,A}$ and $x_{i,B}$ be the reward amounts for turn $i$ of machines A and B, respectively, and let $r_{i,j}$ be the reward amount received by user $j \in 1, N$ in turn $i$. For a given turn $i$, user $j$ makes a choice between A and B, with its reward obeying Rule (ii) such that

$$r_{i,j} = \sum_{k=A,B} \frac{\varepsilon_{i,j,k}}{\sum_{j'=1}^{N} \varepsilon_{i,j',k}} x_{i,k} \;,\; \varepsilon_{i,j,k} \in \{0,1\}. \tag{1}$$

Here, $\varepsilon_{i,j,k}$ corresponds to the actual selection of machine $k$ in turn $i$ by user $j$, with $\sum_{k=A,B} \varepsilon_{i,j,k} = 1$.

After a given number of turns $n_t$, we define $R_j$ and $R$ as

$$R_j(n_t) = \sum_{i=1}^{n_t} r_{i,j} \;;\; R(n_t) = \sum_{j=1}^{N} R_j(n_t). \tag{2}$$

$R_j$ is the accumulated reward for user $j$, and $R$ is the total accumulated reward. Since $N \geq 2$, it is evident that the optimal strategy involves having each machine selected by at least one user in every turn, so that the total accumulated reward $R$ is maximized.

**Fairness**

Optimizing the total accumulated reward $R$ does not imply anything about the repartitioning of the total reward among users. In the simplest case of two users and two machines with constant, different probabilities of giving the same reward, having the users always select the same machines would result in a heavily unbalanced situation. In contrast, if both users recognize the best machine and always select it simultaneously, they receive the same reward amount, although the total accumulated reward $R$ decreases.

To quantify this effect, a metric is needed that indicates in a straightforward manner whether or not the current distribution is fair between users. This notion of fairness has been particularly studied in socioeconomics, in the use of income inequality indices such as the Gini index[27], Hoover index[28], and Theil index[29-31], like the information entropy used in telecommunication. Telecommunication technology also relies on these metrics as figures of merit for resource sharing, with several definitions having been proposed and studied[32-34].

In this study, we decided to use the Jain index $I_J$ to estimate the fairness between users for a given repartition $\{R_j\}_{j \in 1,N}$ after a given number of trials, where $I_J$ is given by

$$I_J = \frac{\left(\sum_{j=1}^{N} R_j\right)^2}{N \cdot \sum_{j=1}^{N} (R_j)^2}. \tag{3}$$

The choice of this metric is justified by several important properties, following the discussion by [34]:

- The metric should be continuous with respect to the reward variables;
- The definition should not vary with *N*, explaining the normalization by *N* in the denominator;
- Its value interval is between 0 and 1;
- It is symmetric and sensitive to the variation in the reward of any user.

The first two properties are essential to establish a metric that is valid for any *N* and any reward distribution. Concerning the Jain index specifically, its main advantage is its intuitiveness: if only one user receives a reward while $N-1$ others do not, then $I_J = \frac{1}{N}$, and for $k$ equally rewarded users and $N-k$ left-out ones, $I_J = \frac{k}{N}$. This aspect explains why $I_J$ is already used widely[34] in telecommunication to estimate the fairness of bandwidth allocation between channels.

Finally, fairness can be understood in two inequivalent ways:

- Fairness of the reward accumulated so far, or
- Fairness of the expected value of rewards to be obtained in the next trial (regardless of the accumulated reward differences).

The first case corresponds to active fairness, in which it is attempted to correct any uneven accumulated repartition in every step. The second case is characterized by a passive pattern, in which it is attempted to make the subsequent trials fair, without trying to correct unfairness *from* past events. In the example of telecommunication, an algorithm following the first case will attempt to compensate for differences in the *accumulated* data amount in every trial, while the second one will only attempt to achieve an equal data *transmission rate* on average between users for the subsequent trials. Since the second situation corresponds to a sufficient condition for the first situation after a transition period, we will focus here only on *passive* or *instantaneous* fairness. Note that the individual rewards in every step do not need to be equal for the situation to be fair: it only matters that the average expected value is the same for all users.

**Targets and performance**

With these tools, we can formulate the problem to be solved in this CMAB situation:

**How can the maximum total reward available for all users be obtained, while simultaneously guaranteeing instantaneous fairness between users?**

Considering this objective, we used a modified version of the Jain index that involves both fairness and total reward performance. Let $X_k(n_t) = \sum_{i=1}^{n_t} x_{i,k}$ be the accumulated reward given by machine $k$ after $n_t$ trials and $X(n_t) = X_A(n_t) + X_B(n_t)$ be the total reward available to users. The modified index is defined as

$$I_p = I_J . X = \frac{\left(\sum_{j=1}^{N} R_j\right)^2}{N . \sum_{j=1}^{N} \left(R_j\right)^2} \times \left(\frac{\sum_{j=1}^{N} R_j}{\sum_{k=A,B} X_k}\right). \tag{4}$$

This quantity corresponds to the fairness index considered by the global performance, which we call the pondered index. It possesses the same advantages as the Jain index while also evaluating the efficiency of the global strategy: if all users make the same choice, the repartition will be fair with poor performance, while a single user who always selects the best machine alone will generate a strongly unbalanced repartition. Thus, $I_p \in [0,1]$ and is equal to 1 if and only if the repartition is fair and the total available reward is obtained by the users. The strategies discussed herein are all intended to make $I_p$ as close to 1 as possible.

***N* = 2 case and constraints on the required quantum state**

This section discusses the *N* = 2 case from the perspective of the fairness discussion introduced before and with the use of quantum superposition states employing the polarization of photons. The purpose is to apply the formalism to a well-known case so that situations with more than two users can be easily treated.

This situation is the simplest case of users having to share limited and uncertain resources among them, without direct communication between them. One possibility is to ensure, by construction, equality among users in the case of a sub-optimal reward produced by a selfish strategy. To maximize the total reward, each machine should be selected by one user in every step, and to maximize fairness, each user should select one machine as often as the other. In other words, users should avoid conflicts of decision between them. This principle leads to the difficult task of guaranteeing that users never select the same machine simultaneously when they cannot communicate with each other.

In our previous work[26], we studied the situation in which users receive photon pairs in a quantum superposition of polarization states, while only being able to rotate a half-waveplate to optimize their rewards. Comparisons with other strategies such as independent random decision making and situations with selfish users are also provided. While the latter part will not be discussed here, we will describe the quantum state used and its properties in relation to the two-user CMAB.

**Derivation of the state**

We demonstrated previously[26] that polarization-entangled photon pairs can solve this issue when they are in state $|\psi(\phi)\rangle_2$ given by

$$|\psi(\phi)\rangle_2 = \frac{1}{\sqrt{2}}\left(|HV\rangle + e^{i\phi}|VH\rangle\right), \qquad (5)$$

where $\phi$ is a real number and $|H\rangle$ and $|V\rangle$ are the horizontal and vertical linear polarization states of single photons, respectively. In that setup, each user has a half waveplate, which transforms any input state according to the angle $\theta/2$ of its fast axis with respect to the $|H\rangle$ direction of the photon source, following

$$\begin{pmatrix} a|H\rangle \\ b|V\rangle \end{pmatrix} \rightarrow \hat{r}(\theta)\begin{pmatrix} a|H\rangle \\ b|V\rangle \end{pmatrix}, \quad \hat{r}(\theta) = \begin{pmatrix} \cos\theta & \sin\theta \\ -\sin\theta & \cos\theta \end{pmatrix}. \qquad (6)$$

Generally speaking, state $|\psi(\phi)\rangle_2$ only works if users have the same polarization measurement base $\{|H\rangle, |V\rangle\}$, or equivalently if $\theta = 0\,[90°]$ with respect to the photon source (there is then no mixing of states while projecting onto the orthogonal states of a polarizing beam splitter cube). In contrast, for $\phi = \pi$,

$$|\psi\rangle_2 = \frac{1}{\sqrt{2}}\left(|HV\rangle - |VH\rangle\right), \qquad (7)$$

which is invariant under any simultaneous rotation of the polarization measurement bases of the users. In this case, there is no need for direct communication between users to achieve an optimal situation, since only the relative angle between the measurement bases of the users matters.

**Properties of the state**

Next, assuming the state $|\psi\rangle_2$ is sent, we discuss the evolution of fairness and total performance of the users when they modify their polarization bases independently by rotating their half-waveplates. It is assumed that this state is well controlled and reproducible and that the detection efficiency is 100%; in practice, limited detection efficiency can be overcome by post-selection or coincidence schemes, at the expense of lower fidelity of the state. Figure 2 shows the evolution of fairness and total performance with the rotation angles of the waveplates of both users after 1000 trials, averaged over 20 repetitions.

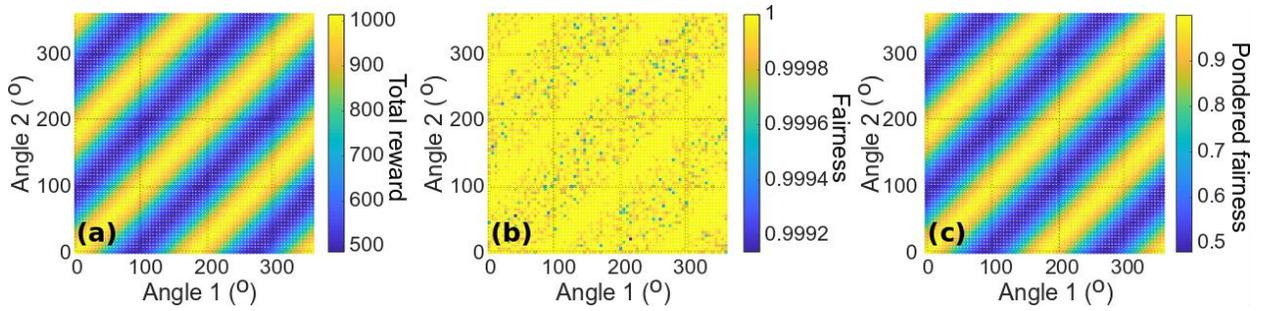

**Figure 2**. **(a)** Total reward, **(b)** fairness, and **(c)** modified Jain index $I_p$ for two users and two machines, using $|\psi\rangle_2$ as the input state, as functions of the measurement basis tilt of each user.

Figure 2(b) shows that the fairness remains constant regardless of the combination of angles, which means that both players always receive the same accumulated reward regardless of their own angles. In addition, Fig. 2(a) demonstrates that the total reward depends only on the relative angle between the waveplates of the users, in accordance with the global rotational invariance of the state. From the perspective of the users, improving the reward of one user is thus equivalent to improving the rewards of both users equally. Finally, Fig. 2(c) presents the modified Jain index $I_p$, which is equivalent here to the evolution of the total reward due to the perfect fairness in every case between users.

**Realignment algorithm convergence**

This section discusses the determination of how users can actually proceed to have their polarization measurement bases aligned with each other (modulo 180°). According to Fig. 2, for any given tilt angle of one user, the other one can find an optimal position by tuning only its own waveplate. In other words, the action of one user is sufficient to find an optimal configuration for both.

In the previous study[26], only one user attempted to find a correct angle configuration in the realignment algorithm, without any information about the angle of the other user. However, this strategy implies agreement regarding which user should act, because if both users act simultaneously they may not find an equilibrium situation.

These observations motivated us to implement a new, simpler, and scalable strategy, which involves selecting a random waveplate position when too many conflict situations are recorded over time. Given that Rules (i), (ii), and (iii) are verified, all users can recognize simultaneously when there is total conflict of decision and thus when realignment is needed, based on the reward amounts they receive (1/$N$ in the case of success and full conflict). The exact algorithm is presented in Section 1 of the Supplementary information.

To compare the performances of the two algorithms, we numerically studied a set of 100 random initial angles for the measurement basis of both users, with any value between 0° and 360° in increments of 5°, from which one player performed either the algorithm corresponding to the autonomous polarization-basis alignment of our previous study[26] (under assumption (II) in that report) or the proposed memory-based random algorithm. Briefly, the first algorithm supposes that no information is available to either user about the measurement basis position of the other user, and only one user adjusts its basis tilt in small increments until conflict is avoided. Each set of initial angles is repeated 20 times, and the averaged $I_p$ is estimated from the angle configuration in several time steps. We used identical parameters for both algorithms, with a memory capacity of eight reward-giving events, a conflict event threshold of 2 before the angle was changed, and an angle increment of 5° for the previous algorithm. These parameters correspond to a trade-off between sensitivity to conflicts and robustness against possible errors.

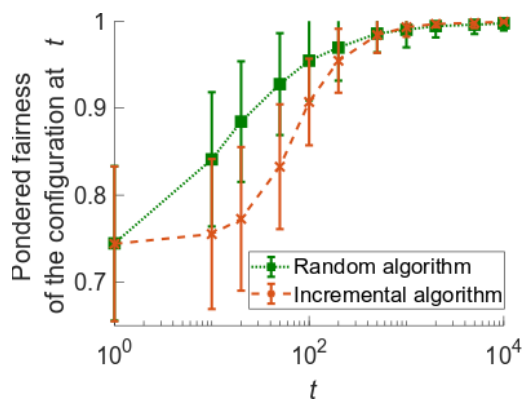

**Figure 3**. Modified Jain index $I_p$ during realignment when one player uses the proposed random algorithm (green line) or the previous incremental algorithm (orange line), for 100 identical random initial angles. Each result is the average of 20 trials.

Figure 3 shows that both algorithms converge toward a stable, optimal configuration after several hundred plays, with the proposed algorithm converging faster. Using a coarser angle increment in the previous algorithm would probably make it converge faster, with the risk of skipping an optimal situation. As such, the proposed algorithm performs well for any initial configuration, and we chose to use it again for realignment estimation for higher numbers of players.

## *N* = 3 case

Based on the considerations in the *N* = 2 case, this section presents the derivation of an appropriate superposition of three-photon polarization quantum states to maximize $I_p$, with a similar ideal setup as in the *N* = 2 case.

**Constraints on the quantum state**

Several criteria on the quantum state to be sent to the users can be introduced following the observations made in the *N* = 2 case:

- Invariance of the measurement probabilities with respect to simultaneous rotation of the measurement bases of all users;
- Symmetry of the state with respect to all other users;
- No term with all users selecting the same choice simultaneously.

The first condition comes from the requirement that the state be device-independent; otherwise, the problem depends on the choice of the measurement basis of the photon source. The second condition means that with an odd number of users, such as three, the terms should have equal probabilities between an unbalanced situation (such as two users on A and one user on B) and its mirror (two users on B and one on A), as well as equal probabilities between all permutations of these terms. Finally, the last condition is associated with the search for the maximum total reward, which requires that all choices be attempted in every trial.

**Derivation of the quantum state**

Let us start from the three conditions given in the previous section to discuss whether or not there exists a family of compatible states for *N* = 3. This derivation is intended to demonstrate how these conditions constrain the form of the final quantum state obtained.

Global rotational invariance is best described using the density matrix formulation for the quantum states. Let $|\psi\rangle_3$ be the entangled state under study and $\rho_3$ the associated density matrix. Following the last criteria given in the previous section and using the polarization basis $\{|H\rangle, |V\rangle\}$ of the photon source, the state in the three-photon Hilbert space can be written as

$$|\psi\rangle_3 = 0|HHH\rangle + a_1|HHV\rangle + a_2|HVH\rangle + a_3|VHH\rangle \\ + b_1|VVH\rangle + b_2|VHV\rangle + b_3|HVV\rangle + 0|VVV\rangle \quad (8)$$

in its canonical basis, where the coefficients are all complex and include the normalization condition. From this equation, the necessary conditions on the coefficients can be derived such that the invariance of the state probability measurements does not vary under a simultaneous rotation $\theta \in \mathbb{R}$ of all measurement bases, denoted hereafter as $\{|\theta\rangle, |\theta'\rangle\}$ in the photon source reference frame. This condition involves the global rotation operator $\hat{R}_3(\theta)$, defined in the three-photon Hilbert space $H$ from the two-dimensional operator $\hat{r}(\theta)$ as

$$\hat{R}_3(\theta) = \hat{r}(\theta) \otimes \hat{r}(\theta) \otimes \hat{r}(\theta), \quad \hat{r}(\theta) = \begin{pmatrix} \cos\theta & \sin\theta \\ -\sin\theta & \cos\theta \end{pmatrix}. \tag{9}$$

Then, the condition of invariance of the measurement probabilities under global rotation can be expressed as

$$\forall \theta \in \mathbb{R}, \forall |\phi\rangle \in H, \langle\phi|\hat{R}(\theta)^\dagger \rho_3 \hat{R}_3(\theta)|\phi\rangle = \langle\phi|\rho_3|\phi\rangle. \tag{10}$$

After applying (10) to the canonical basis of the three-photon Hilbert space, the following non-redundant conditions can be obtained:

$$a_1 + a_2 + a_3 = 0 \tag{11}$$

$$a_1 = \pm ib_1; a_2 = \pm ib_2; a_3 = \pm ib_3 \tag{12}$$

$$|a_1| = |a_2| = |a_3| = |b_1| = |b_2| = |b_3| = \frac{1}{\sqrt{6}}. \tag{13}$$

From (11) and (13), one can deduce that the complex cubic roots of 1 are involved. In addition, as the global phase of the state is irrelevant, there are 11 unknown variables to find (the amplitudes and phases of the parameters), with only 10 equations ((13) summarizes six equations). Four states can solve the equations, corresponding to the permutation of the phase differences and the choice of $\pm i$ in (12):

$$|\psi\rangle_3 = \frac{1}{\sqrt{6}}\left[\left(|HHV\rangle + z|HVH\rangle + z^2|VHH\rangle\right) \pm i\left(|VVH\rangle + z|VHV\rangle + z^2|HVV\rangle\right)\right], z = e^{\pm\frac{2i\pi}{3}}. \tag{14}$$

All states have the same properties that are relevant to the present analysis, with symmetry in the rotation parameter space of each player. In the following, we discuss only the state

$$|\psi\rangle_3 = \frac{1}{\sqrt{6}}\left[|HHV\rangle + i|VVH\rangle + e^{\frac{2i\pi}{3}}\left(|HVH\rangle + i|VHV\rangle\right) + e^{\frac{4i\pi}{3}}\left(|VHH\rangle + i|HVV\rangle\right)\right]. \tag{15}$$

**State properties**

As discussed about the constraints, the total reward for all users is maximized when at least one user selects each machine in every turn. Thus, the total reward is reduced in conflict situations in which all users select one machine: for clarity, we will use the so-called conflict rate as a metric to highlight the loss of global performance rather than total reward.

Figure 4 shows three-dimensional scatter plots of the total reward, fairness, and pondered index as functions of the angle of the polarization measurement basis of each user relative to the photon source basis, with 5° discretization. The color and size of each sphere are functions of the variable studied. The fairness exhibits the greatest difference from the *N* = 2 case here, as there exist situations in which both total reward and fairness are sub-optimal. The total rewards still vary considerably over the parameter space, and the combination of the variations explains the more significant variation of the pondered index with respect to the *N* = 2 case.

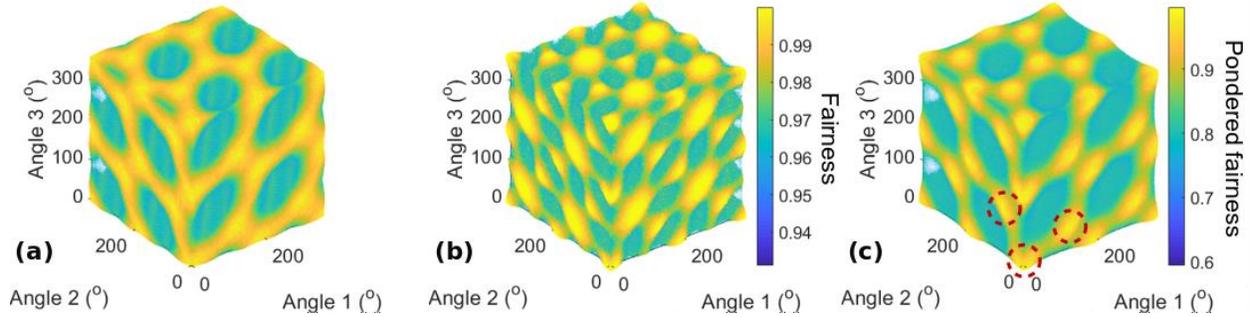

**Figure 4**. **(a)** Total reward, **(b)** fairness, and **(c)** pondered fairness $I_p$ for $N=3$ users rotating their polarization measurement bases independently. Red circles in **(c)** indicate optimal angle combinations from which all other can be deduced by global rotation or a period of $\pi$.

Note that in addition to $\pi$-periodic optimal situations and invariance with respect to simultaneous rotation, additional optimal situations can be found for specific combinations of angles, indicated by the red circles in Fig. 4(c) and corresponding to rotations of $\pi/3$ or $2\pi/3$. This characteristic corresponds to the fact that permutations of the phases of several terms in $|\psi\rangle_3$ yield the same performance. This feature leads to a parameter space in which many combinations of angles between the three users are favorable.

**Realignment algorithm convergence**

In accordance with the objective of achieving device-independent behavior, the users should be able to find the optimal situation without needing to communicate with each other. Supposing that their measurement bases are misaligned with each other, there should be a mean of retrieving a favorable situation as quickly as possible. However, unlike in the $N$ = 2 case, the actions of one user alone are not sufficient to reach the optimal angle combination in the general case, as can be seen in Fig. 4 along the tuning parameter axis of each user taken separately. Consequently, the random exploration strategy introduced in Realignment algorithm convergence of the $N$ = 2 case is needed.

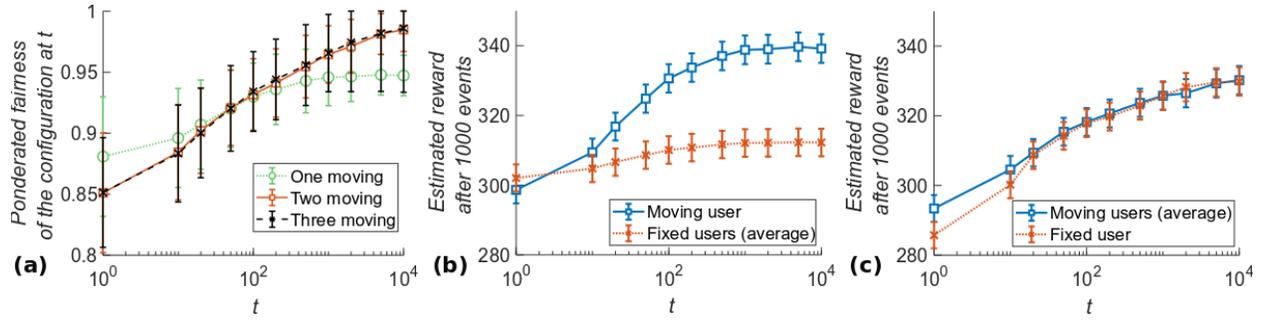

**Figure 5**. **(a)** Instantaneous pondered equality $I_p$ after different numbers of time steps *t* (log scale), when one (green), two (blue), or three (black) users react to misalignment. Estimated upcoming rewards for **(b)** one moving user (blue) and two fixed users (red) and **(c)** two moving users (blue) and one fixed user (red).

To estimate the efficiency and speed of convergence of this strategy, we selected 100 initial random angle combinations and repeated the calculations 20 times for each combination to obtain a statistical average. At different time steps, the angle combination was taken and attempted 1000 times to estimate the instantaneous $I_p$. Figure 5(a) shows the evolution of $I_p$ over time when one, two, or three users follow this strategy while the other(s) remain fixed, still with 5° discretization of the available angles for computational purposes. It is evident that when at least two users react to conflicts, the average $I_p$ increases steadily until it reaches more than 0.98 after 10,000 events, while it remains bounded at 0.95 on average when only one user moves. At least two users are thus necessary to reach the optimal situation.

We also tested the evolutionary stability [Taylor1978] of this strategy by checking whether users are at a disadvantage or an advantage if they use this strategy or if they remain passive, relying on other users to improve the situation. Figure 5(b) depicts the estimated reward given the current angle configuration for each user when only user 1 is moving, while Fig. 5(c) shows the situation when users 1 and 2 react to conflicts, but not user 3. As can be seen, a passive user always receives a lesser or equal reward compared to the active users, while all users always benefit from attempting to improve the situation. It is thus beneficial for users to follow this algorithm rather than remaining passive, which confirms the evolutionary stability of the algorithm. Besides, if any user tries to increase its own outcome, other users can easily recognize it and get even higher reward than the selfish user, thus pushing users to play for the common best situation.

## *N* = 4 case

We have already identified the differences between odd and even numbers of users, mainly the necessary symmetry between terms that gives unbalanced rewards, such as when there are two users for machine A and one user for machine B, as well as similarity, such as invariance by simultaneous rotation of the measurement bases of the users. This part will focus on the $N = 4$ case to identify the properties that are preserved when the situation is scaled up according to the number of users.

**Differences from the *N* = 2 and *N* = 3 cases**

The same constraints apply to the state to be obtained with respect to the $N$ = 3 case: invariance with respect to simultaneous rotation of the polarization measurement bases, symmetry between users, and avoidance of terms in which all users select the same machine simultaneously. However, this last constraint is fulfilled by two types of terms:

- Asymmetric terms in which three users select one machine and one user selects the other, such as $|HHHV\rangle$ and $|VVVH\rangle$,
- Symmetric terms in which users are split equally between machines, such as $|HHVV\rangle$ and its permutations.

A priori, both kinds of terms are usable here as long as their relative amplitudes are equal between permutations of the photons of the users; other considerations are required to determine the expression of the state.

**Derivation of the quantum state**

The state can be derived using the same technique as for *N* = 3, starting from global rotation invariance of the density matrix of the target state. We chose to study separately the asymmetric and symmetric terms for clarity.

Let $|S\rangle_4$ and $|A\rangle_4$ respectively be the asymmetric and symmetric target states in the corresponding Hilbert space, with all coefficients being complex:

$$|S\rangle_4 = c_1|HHVV\rangle + c_2|HVHV\rangle + c_3|HVVH\rangle + c_4|VHHV\rangle + c_5|VHVH\rangle + c_6|VVHH\rangle \quad (16)$$

$$\begin{aligned}|A\rangle_4 = &a_1|HHHV\rangle + a_2|HHVH\rangle + a_3|HVHH\rangle + a_4|VHHH\rangle \\ &+ b_1|VVVH\rangle + b_2|VVHV\rangle + b_3|VHVV\rangle + b_4|HVVV\rangle\end{aligned} \quad (17)$$

Then, the invariance by simultaneous rotation of any angle $\theta$ using the operator $\hat{R}_4(\theta)$ is applied. For the symmetric states $|S\rangle_4$, the conditions on the complex coefficients for invariance under operator $\hat{R}_4(\theta)$ follow:

$$c_1 + c_2 + c_3 = 0 \quad (18)$$

$$c_1 = c_6, \quad c_2 = c_5, \quad c_3 = c_4 \quad (19)$$

$$|c_1| = |c_2| = |c_3| = |c_4| = |c_5| = |c_6| = \frac{1}{\sqrt{6}}. \tag{20}$$

As before, the global phase can be fixed at 0 without loss of generality, which gives 10 equations for 11 unknowns. The constraints are the same as for the *N* = 3 case, with the difference that there is no $\pm i$ factor anymore between mirror terms. There are two possible states of this kind:

$$|S\rangle_4 = \frac{1}{\sqrt{6}}\left(|HHVV\rangle + |VVHH\rangle + z(|HVHV\rangle + |VHVH\rangle) + z^2(|HVVH\rangle + |VHHV\rangle)\right), z = e^{\pm\frac{2i\pi}{3}}. \tag{21}$$

In the following, we will discuss only the case $z = e^{\frac{2i\pi}{3}}$, as both states produce the same results in this analysis.

For $|A\rangle_4$, the global rotation invariance gives the conditions

$$a_1 + a_2 + a_3 + a_4 = 0 \tag{22}$$

$$a_1 = -b_4, \quad a_2 = -b_3, \quad a_3 = -b_2, \quad a_4 = -b_1 \tag{23}$$

$$|a_1| = |a_2| = |a_3| = |a_4| = |b_1| = |b_2| = |b_3| = |b_4| = \frac{1}{\sqrt{8}}. \tag{24}$$

Excluding the global phase, there are 15 unknowns for 13 equations, leaving the choice of one relative phase between $a$ terms and the permutation order (for a given phase difference between, say, $a_1$ and $a_2$, we can have either $a_3 = -a_1$ or $a_4 = -a_1$). This characteristic gives the family of states

$$|A\rangle_4(\phi) = \frac{1}{\sqrt{8}}\Big[(|HHHV\rangle - |VVVH\rangle) + e^{i\phi}(|HHVH\rangle - |VVHV\rangle) \\ -e^{i\frac{\phi}{2}}\Big[e^{\pm i\frac{\phi}{2}}(|HVHH\rangle - |VHVV\rangle) + e^{\mp i\frac{\phi}{2}}(|VHHH\rangle - |HVVV\rangle)\Big]\Big], \tag{25}$$

where $\phi \in [0, 2\pi]$. At this stage, the choice of $\phi$ is not constrained, as all these states follow all the rules we have set.

**State properties**

As before, the objective is to study the variations of the performances of the states when the users vary their waveplate angles. As a four-dimensional representation of the parameter space is inconvenient to display and comprehend, we used the global rotation invariance and fixed the angle of one user at 0°, while the three other users varied their angles. The other configurations can be deduced by translation of the three-dimensional figure along the three-dimensional diagonal.

Figure 6 shows the total reward, fairness, and $I_p$ for the symmetric state $|S\rangle_4$ when users 1, 2, and 3 tilt their waveplates, with 5° discretization, and user 4 remains fixed at 0. While the fairness remains almost optimal in any configuration, the conflict rate (and, thus, the total reward) changes significantly while only particular states give optimal situations, typically with a period of 180°. Between them, conflicts and/or fairness are sub-optimal, and the density of optimal combinations in the four-dimensional space is relatively low. Regarding realignment, at least three users generally need to tune their half-waveplates to achieve an optimal situation. Figure 6(c) shows that for any fixed angle, for example, for user 3, there is no optimal configuration unless users 3 and 4 have the same angle (here, 0°). This finding is confirmed by the simulation results presented in Section 2 of the Supplementary information.

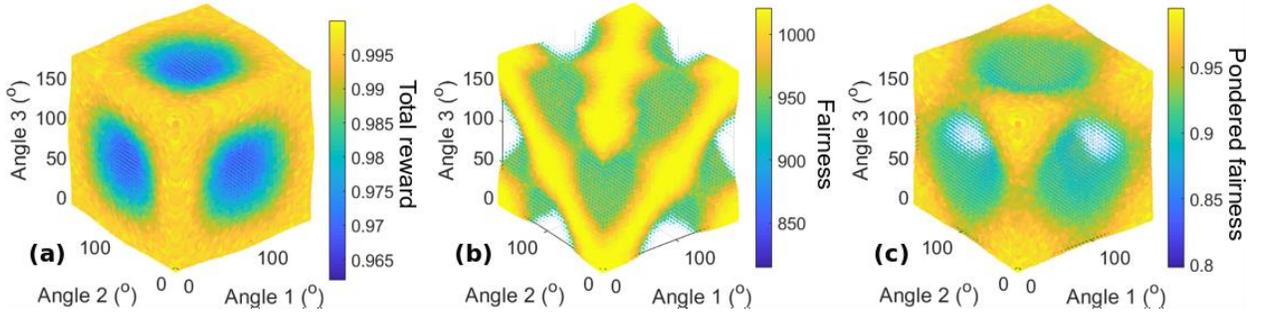

**Figure 6.** **(a)** Total reward, **(b)** fairness, and **(c)** pondered fairness $I_p$ for N = 4 users and $|S\rangle_4$, when three users tune their waveplates and the fourth remains static at 0°.

All of the asymmetric states $|A\rangle_4(\phi)$ show fairness greater than 0.995 for any angle configuration. Thus, the variations of $I_p$ only depend on the total reward. To understand the influence of $\phi$ on the state performance, we studied the following states:

$$|A\rangle_4(\phi) = \frac{1}{\sqrt{8}}\Big[\big(|HHHV\rangle - |VVVH\rangle\big) + e^{i\phi}\big(|HHVH\rangle - |VVHV\rangle\big) \\ -\big[\big(|HVHH\rangle - |VHVV\rangle\big) + e^{i\phi}\big(|VHHH\rangle - |HVVV\rangle\big)\big]\Big] \qquad (26)$$

Figures 7(a), (b), and (c) show only $I_p$ for the asymmetric states $|A\rangle_4(0)$, $|A\rangle_4\left(+\frac{\pi}{2}\right)$, and $|A\rangle_4(\pi)$, respectively, with the same user configuration as for $|S\rangle_4$. For $\phi = 0$ and $\phi = \pi$, there exist several planes of optimal situations, passing through points $(0[\pi], 0[\pi], 0[\pi])$ that are not on the same edge of the cube of length $\pi$. However, for $\phi = \frac{\pi}{2}$, the optimal situations are only on lines corresponding to the intersections of the planes visible in (a) and (c).

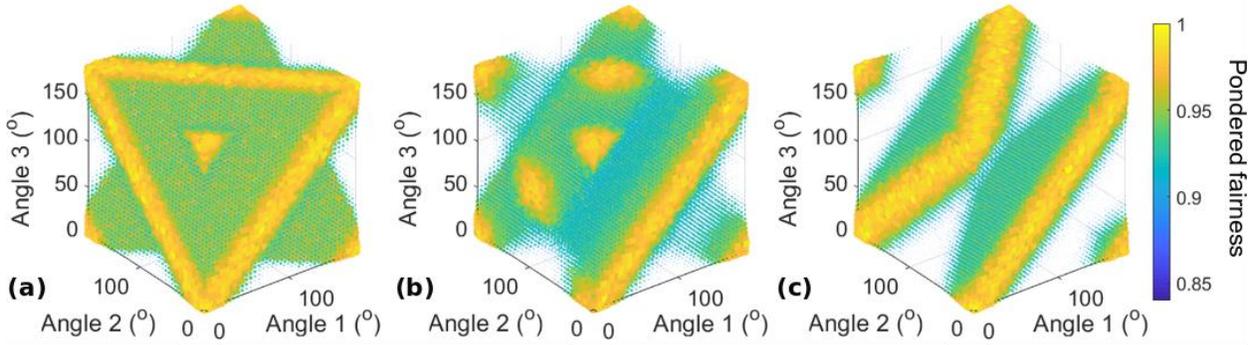

**Figure 7**. Pondered fairness $I_p$ for $N=4$ users and $|A\rangle_4(\phi)$ for different values of $\phi$, when three users tune their waveplates and the fourth remains static at 0°. **(a)** $\phi=0$, **(b)** $\phi=\dfrac{\pi}{2}$, and **(c)** $\phi=\pi$. The fairness $I_J$ is maximized within the natural fluctuation level for all these configurations.

Optimal planes such as those in Figs. 7 can be obtained for any permutation of the terms in (26) as long as $\phi=0[\pi]$. The locations of these planes can be explained by the choice of phase terms in (26) and the sign of the phase difference in the second group of terms. Different signs give different planes in the parameter, although they always pass through points $(0[\pi],0[\pi],0[\pi])$ that are not on the same edge of the cube of length $\pi$ for a given plane. Furthermore, if user 4 has any fixed angle $\theta_4$, the pattern in the parameter space of the three remaining users will simply be shifted, and the new optimal planes will pass through points $(\theta_4[\pi],\theta_4[\pi],\theta_4[\pi])$.

Regarding the realignment procedure, in all cases and for any combination of fixed angles for two users, such as users 3 and 4, there always exists an optimal situation accessible to the two remaining users, as can be seen in Fig. 7. Thus, two users are sufficient to obtain optimal conditions, even though the other users remain fixed. In addition, fairness is always achieved; thus, it is of the best interest of each user to improve the situation for everyone, which is directly linked to the situation of the given user. This finding is also confirmed by the simulation results corresponding to this case, as presented in Section 2 of the Supplementary information. Besides, when $\phi=0[\pi]$, one user can reach an optimal situation from any initial 4-angle configuration by itself, as all lines parallel to x, y or z axes intersect an optimal plane: in those particular cases, the action of one user is enough to realign the whole system.

## DISCUSSION

In the previous sections, we demonstrated the application of quantum superposition of the polarization states of two, three, and four users who must decide between two machines. Explicit expressions of the optimal states in each case were found, as well as expressions of the optimal states in the $N = 5$ case, whose derivation is presented in Section 3 of the Supplementary information. In addition, we developed a script using Wolfram Mathematica, presented in Section 4 of the Supplementary information, that can identify whether or not a given $N$-photon state is symmetric and invariant under simultaneous rotation of all users. A generalization of the theoretical work presented here for any $N$ would be an interesting development, as entangled photon states with more than 10 photons have already been generated experimentally. Based on the empirical observations, we hypothesize that complex $N$-roots of 1 are involved in the phase coefficients for the optimal state formulation, regardless of $N$.

In addition, we observed differences between odd and even $N$: while maximum fairness is achieved for even $N$ regardless of the combination of user waveplate angles, odd $N$ does not exhibit this property. This fact remains to be explained.

As for any resource sharing system, security is a major concern similarly with all telecommunication systems. The system described in this article relies on two fundamental features:

- The properties are device independent, so any new user can be added with initial alignment verification.
- Equality among users and resource allocation performance can be achieved without having to rely blindly on a remote central entity.

Regarding the first point, device independence is guaranteed by the rotation-independent property of the quantum states used in this study. Regarding the second point, the central entity corresponds to the $N$-photon source providing for all users. For this objective to be achieved, the users need a way to check whether the state received indeed corresponds to the optimal state expected (given the total number of simultaneous users and the expected total throughput). Such protocols have already been developed and verified for entangled photon pairs in the case of quantum key distribution[36-38], including in a star-type network between an arbitrary number of users[39], and several studies have proven the feasibility for more than two parties[40-42], although limitations have been identified for large $N$[43]. All this may be of particular interest to secure applications of quantum information and resource sharing such as voting strategies using quantum states[44].

Last but not least, this analysis relies on means of producing fully tunable quantum superposition of $N$ photons with their polarizations as the entangled degree of freedom. Recent works have shown ways to obtain a Greenberger–Horne–Zeilinger state for up to 10 photons[45], while experiments have already succeeded in generating any 6-photon state[46]. As such, the production of $N$-photon states is less a theoretical issue than a technological one.

**CONCLUSION**

Solving the CMAB problem implies maximizing the total outcome from given choices, as well as ensuring fair repartition of this outcome among all users. In this article, we theoretically and numerically demonstrated that carefully chosen polarization-entangled $N$-photon states can solve this problem for at least five players who must choose simultaneously between two choices or machines. Different behaviors, such as guaranteed fairness of outcome repartition only for even $N$, were identified between even and odd numbers of players, due to the fundamental properties of the polarization degree of freedom of photons. Nonetheless, in every case, the properties of these states imply that the best strategy for each player is to attempt actively to reach a measurement configuration that corresponds to a common optimum for all players simultaneously, thus achieving evolutionary stability. These quantum states are derived from a few basic rules, such as global invariance by rotation of the measurement bases of the players and symmetry of the state by permutation of the players and/or choices. Although a detailed derivation of the favorable states for any $N$ has not been performed, we developed an algorithm that can verify whether any entangled state follows the set of rules described in this report. In addition, we demonstrated that, under basic assumptions, a fairly simple one-sided algorithm enables the users to correct any misalignment from an optimal configuration autonomously, without the need to trust a central entity. By associating this kind of system with Bell-type verification of the $N$-photon-state source, decentralized, secure, and optimal resource sharing among users could be achieved.

**Supplementary information**

1. Realignment algorithm for any *N*

2. Realignment for *N* = 4

3. State for *N* = 5

4. Mathematica script for optimal state verification

**Data availability**

The datasets generated during the current study are available from the corresponding author on reasonable request.

## Acknowledgements

This work was supported in part by the CREST project (JPMJCR17N2) funded by the Japan Science and Technology Agency and Grants-in-Aid for Scientific Research (JP17H01277 and JP20H00233) funded by the Japan Society for the Promotion of Science.

## Author Contributions

M.N., G.B., S.H. and N.C. directed the project. H.S. and H.H. provided theoretical fundamentals and insight. N.C. developed the simulations and the Mathematica script. M.N., G.B., S.H. and N.C. analyzed the data. M.N., G.B., S.H. and N.C. wrote the paper.

## Additional information

**Competing interests:** The authors declare no competing interests.

**Correspondence and requests for materials** should be addressed to N.C.